
\input mtexsis
\texsis
\preprint
\input epsf.tex

\pubcode{PSU/TH/114}

\titlepage
\title
On $\pi -\pi $ Correlations in Polarized Quark Fragmentation Using the
Linear Sigma Model
\endtitle
\author
John C. Collins
Department of Physics
Penn State University
104 Davey Laboratory
University Park, PA  16802, U.S.A.
\endauthor
\and
\author
Glenn A. Ladinsky
Department of Physics and Astronomy
Physics--Astronomy Building
Michigan State University
East Lansing, MI  48824, U.S.A.
\endauthor
\abstract
Using the linear sigma model to describe quark--pion
interactions, we compute
polarization asymmetries in quark fragmentation.
We show that the effects
of transverse quark polarizations appear
in the correlation between the two leading pions in a jet produced by the
fragmentation of a quark.  Such asymmetries provide a window to the
nature of chiral symmetry breaking in QCD.
\endabstract
\endtitlepage
\def\mynot#1{\not{}{\mskip-3.5mu}#1 }
\def\del{\partial }
\def\half{{1\over 2}}
\def\vpi{\vec{\bf\pi }}
\def\kt{{\bf k}_{T}}

\def\journal#1&#2(#3)#4{{\unskip,~\sl #1\unskip~\bf\ignorespaces #2\unskip~\rm
(19#3) #4}}

\def\gee{{\tilde{g}}}

\referencelist
\reference{kly}
G.L.~Kane, G.A.~Ladinsky and C.-P.~Yuan \journal Phys.~Rev. &D45 (92) 124
\endreference
\reference{kayu}
G.L.~Kane and C.--P.~Yuan \journal Phys.~Rev.&D40 (89) 2231
\endreference
\reference{jccgal}
J.C.~Collins and G.A.~Ladinsky, in
{\it Proceedings of the Polarized Collider Workshop}, Penn
State University, 1990, ed.\ J.C.~Collins, S.~Heppelmann and R.~Robinett
(AIP Conference Proceedings No.\ 223, American Institute of Physics,
New York, 1991), pg. 215
\endreference
\reference{nacht}
O. Nachtmann \journal Nucl.~Phys.\ &B127 (77) 314
\endreference
\reference{umconf}
G.R.~Goldstein, in ``Determining Quark Helicity from Jet Distributions''
in Proc.\ of 8th Int.\ Symp.\ on High Energy Spin Physics, Minneapolis, MN,
(e.d, K.J.~Heller) A.I.P. Conference Proceedings \#187
(A.I.P., New York, 1989), pg. 291
\endreference
\reference{*umconf}
M.B.~Einhorn, in ``Spin Correlations in Quark and Gluon Fragmentation''
in Proc.\ of 8th Int.\ Symp.\ on High Energy Spin Physics, Minneapolis, MN,
(ed.,\ K.J.~Heller) A.I.P. Conference Proceedings \#187
(A.I.P., New York, 1989), pg. 802
\endreference
\reference{efremov}
A.V.~Efremov, L.~Mankiewicz and N.A.~T\"ornqvist \journal
Phys.~Lett.&B284 (92) 394
\endreference
\reference{gang}
J.C.~Collins, S.~Heppelmann, and
G.~Ladinsky\journal Nucl.~Phys.&B420 (94) 565
\endreference
\reference{ccjpap}
J.C. Collins\journal Nucl.~Phys.&B396 (93) 161
\endreference
\reference{georgi}
A.~Manohar and H.~Georgi \journal Nucl.~Phys.&B234 (84) 189
\endreference
\reference{gellmann}
M.~Gell--Mann and M.~Levy\journal Nuovo~Cimento&16 (60) 705
\endreference
\reference{*chengli}
also see, for example,
T.-P. Cheng and L.-F. Li,
{\it Gauge Theory of Elementary Particle Physics}, Clarendon Press, Oxford,
England, 1984
\endreference
\reference{sparam}
W. Lin and B.D. Serot \journal Nucl.~Phys.&A512 (90) 637
\endreference
\reference{*sparamb}
T.~Schr\"oder \journal Europhys.~Lett.&12 (90) 497
\endreference
\reference{jcpdf}
J.C.~Collins and D.E.~Soper \journal Nucl.~Phys.&B194 (82) 445
\endreference
\reference{swidtha}
M.~Nielsen and J.~da~Provid\^encia \journal Phys.~Lett. &B271 (91) 21
\endreference
\reference{swidthb}
M.D.~Scadron \journal Mod.~Phys.~Lett.&A7 (92) 497
\endreference
\endreferencelist

\section{Introduction}

Measuring the polarization of final state particles has been of
continued interest in high energy particle physics.  Such observations
facilitate the measurement of the form factors which govern particle
interactions and provide a probe of the symmetry breaking mechanisms
present in nature.\cite{kly}\cite{kayu}\cite{jccgal}
One interesting idea is to probe the polarization carried by
a quark or gluon of QCD through the distribution of its
fragmentation products in a final state
jet, first proposed by Nachtmann.\cite{nacht}\cite{umconf}  In particular,
Nachtmann\cite{nacht} showed that a three particle
correlation in a jet can probe the helicity of the parton
initiating the jet.  This idea was rediscovered by Efremov,
Mankiewicz and T\"ornqvist\cite{efremov}, and they named the
correlation the `handedness' of a jet.

In collaboration with Heppelmann\cite{gang},
we showed that, in a
similar fashion, a {\it two-particle} correlation can be used to
probe the {\it transverse} polarization of the quark initiating a
jet.
This is a novel idea, and there is at present no known
experimental information on this two-particle correlation.
In this paper, we will perform some very simple model
calculations of the transverse spin dependence of
fragmentation.  We will use a linear sigma model of pions
coupled to quarks, somewhat in the spirit of Georgi and
Manohar.\cite{georgi}  We will assume that this model, taken in
the lowest relevant order of perturbation theory,
is an
approximation to the long distance dynamics of QCD.
Within the model, we will calculate the spin-dependent part
of the fragmentation of quarks to a two-pion state,
$q\to \pi \pi +X$.
The spin dependence shows up as a correlation between the plane
of the two-pion system and the transverse spin vector of the
quark.

The importance of our calculation is that it shows that
there is consistency between the
symmetries of QCD and
a nonzero transverse-spin-dependence for fragmentation at the
leading-twist level.  Indeed, the large effect we calculate shows
that there is no suppression of the spin-dependence.
The spontaneous breaking of chiral symmetry
(and hence a non-zero pion mass)
is essential to our calculation.  Moreover some
kind of nontrivial phase or interference is also essential.
Since we are in a strong-coupling regime, this does not
preclude a large effect.
Indeed, the size of the effect is very large---$50\%$ or more.
In our simple model, the
interference is between two-pion production in the continuum
and at the $\sigma $ resonance.

After reviewing the sigma model and the methods for computing decay
functions, unpolarized quark fragmentation to a $\sigma $ is
calculated.  This is used to normalize the polarization
asymmetry.  We then
compute the polarized decay function to a two-pion state, and
establish a nonvanishing asymmetry.  This then demonstrates how the
correlation between the two pions reflects the spin of the quark.

Finally we present some numerical calculations.

\section{Sigma Model, Feynman Rules and Decay Functions}

\subsection{Sigma Model Lagrangian}

We modify the sigma model lagrangian\cite{gellmann} to use
quarks instead of nucleons:
$$
{\cal L}=
\half [(\del_{\mu }\sigma )^{2}+(\del_{\mu }\vpi )^{2}]+
                i\bar{q}\gamma ^{\mu }\del_{\mu }q
-\gee\bar{q}(\sigma +i\vec{\bf\tau }\cdot
            \vpi \gamma _{5})q-V(\sigma ^{2} +\vpi ^{2})
\EQN{lag}
$$
where
$$
V=-{\mu ^{2}\over 2}(\sigma ^{2} +\vpi ^{2})+
            {\lambda \over 4}(\sigma ^{2} +\vpi ^{2})^{2}.
$$
The Lagrangian possesses an $SU_{R}(2)\times SU_{L}(2)$ symmetry, and
the term linear in the $\sigma $ field that is normally used to
provide explicit breaking of the chiral symmetry is not
relevant for our purposes.

The sigma field is an isosinglet while the pion field is an
isovector whose components may be written
$$
\vpi =\left(\matrix{\pi _{1}\cr\pi _{2}\cr\pi _{3}\cr}\right).
\EQN{pivec}
$$
By using the isospin invariance of the pion field we can define the charged
and neutral pion fields,
$$
\pi ^{\pm }\equiv {1\over\sqrt {2}}(\pi _{1}\mp i\pi _{2}),
                    \qquad\pi ^{0}\equiv \pi _{3}.
\EQN{newpis}
$$
The quark field is an isodoublet with three colors,
$$
q =\left(\matrix{u\cr d\cr}\right).
\EQN{qfield}
$$
The $(\tau ^{a})_{ij}$ are the $2\times 2$ Pauli matrices,
$$
\tau ^{1}=\left(\matrix{0 & 1 \cr 1 & 0 \cr}\right),
\qquad
\tau ^{2}=\left(\matrix{0 & -i \cr i & 0 \cr}\right),
\qquad
\tau ^{3}=\left(\matrix{1 & 0 \cr 0 & -1 \cr}\right).
\EQN{pauli}
$$
We use an index $a=1,2,3$ for the pion fields and indices
$i,j=1,2$ for the
quark fields.

Because of the wrong sign mass term in \Eq{lag}, there is
spontaneous symmetry breaking.  We choose the $\sigma $ field to have the
nonzero vacuum expectation value:
$$
\langle 0|\sigma |0\rangle =v,
    \qquad \langle 0|\vpi |0\rangle =0, \qquad v=\sqrt {\mu ^{2} /\lambda }.
\EQN{brake}
$$
After the transformation $\sigma \rightarrow \sigma +v$,
the quark interaction becomes
$$
-\gee\bar{q}(\sigma +i\vec{\bf\tau }\cdot \vpi \gamma _{5})q
\longrightarrow
-\gee v\bar{q}q-\gee\bar{q}(\sigma +i\vec{\bf\tau }\cdot \vpi \gamma _{5})q ,
\EQN{newqq}
$$
and the remaining $\vec{\bf\pi }$ and $\sigma $ interactions become
$$
-V(\sigma ^{2} +\vpi ^{2})\longrightarrow -\mu ^{2}\sigma ^{2}
-\lambda v\sigma ^{3}
-{\lambda \over 4}\sigma ^{4} -{\lambda \over 4}\vpi ^{4}
-{\lambda \over 2}\sigma ^{2}\vpi ^{2}
-\lambda v\sigma \vpi ^{2}  .
\EQN{newv}
$$
Note that the pion field is massless,
the $\sigma $ has a mass of $\sqrt {2}\mu $
and the quarks have gained a mass of $\gee v$.  If we take the up quark mass
to be around $300\,$MeV and the vacuum expectation value of the sigma field
for the spontaneous symmetry breaking to be $v=f_{\pi }=92\,$MeV, then
$\gee\approx 3$.  Typically, the sigma mass is taken to be
around\cite{sparam} $600\,$MeV.
The couplings $\gee$ and $\lambda $ are large, as is appropriate for
hadronic interactions in the non-perturbative regime.

\subsection{Feynman Rules and Decay Functions}

Among the Feynman rules obtained from the Sigma Model lagrangian after
spontaneous symmetry breaking are the vertex factors, as
displayed in \Fig{ftwo}.
To perform a calculation involving charged pions it is useful
to introduce isospin ``polarization'' vectors:
\def\veps{\varepsilon}
$$
\veps^{a}_{\mp }= (1,\pm i,0)/\sqrt {2},
\qquad
\veps^{a}_{0}= (0,0,1).
\EQN{polvecs}
$$
An incoming state gets an $\vec{\bf\veps}$ and an outgoing state gets an
$\vec{\bf\veps}^{\,*}$.  A color average over the initial state is used.

\figure{ftwo}
\epsfysize=5in
\centerline{\epsfbox{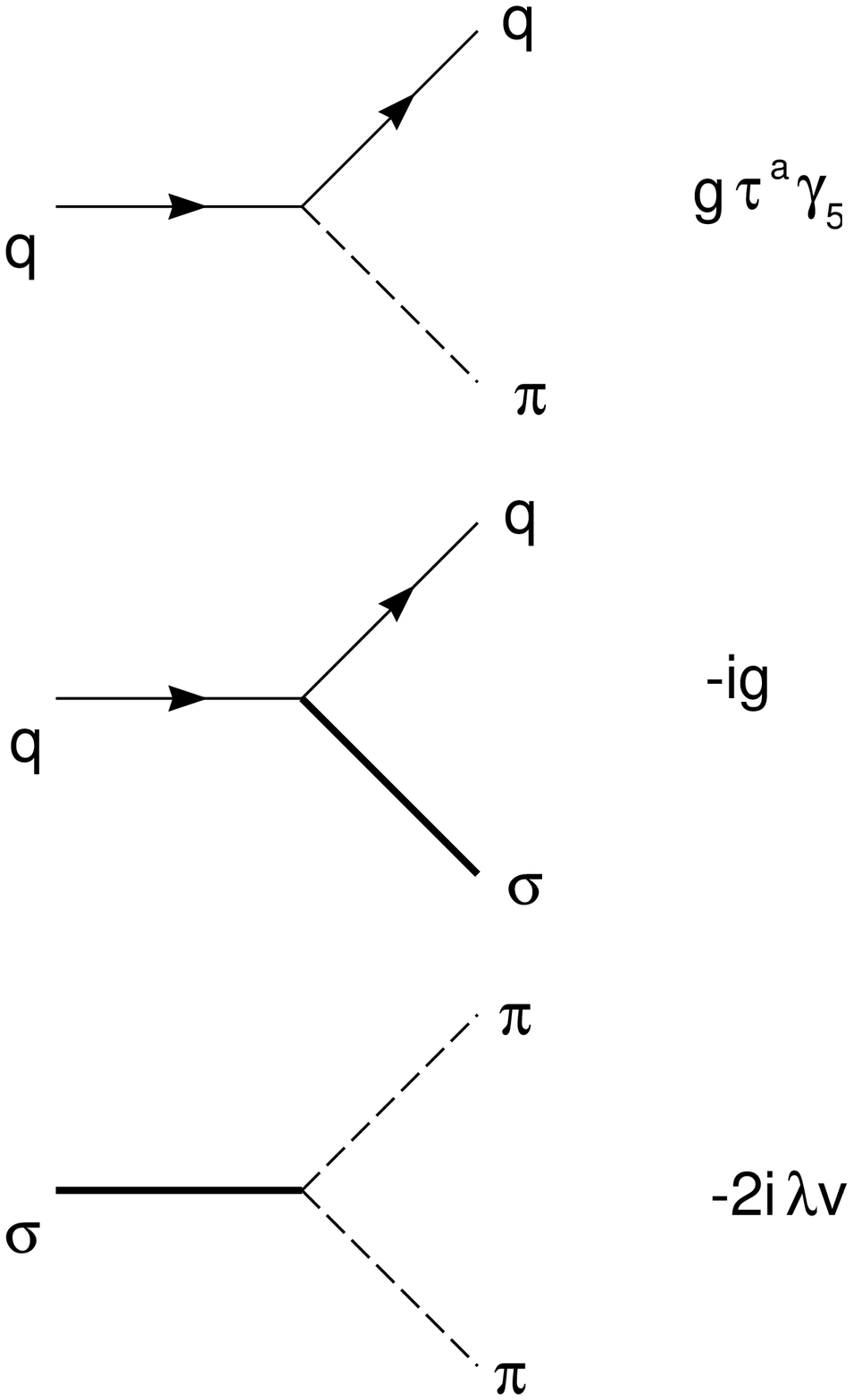}}
\caption{The relevant Feynman rules obtained from the Sigma Model lagrangian.}
\endfigure

In our model, we will compute an approximation to the
spin-dependent fragmentation of a quark with helicity $h$
and space-like transverse spin vector $s_{T}^{\mu }$.
The total spin will obey the
constraint $0\le h^{2}-s_{T}^{2} \le 1$.  (Note that with the usual
metric, $s_{T}^{2} \le 0$.)
In the definition of the
fragmentation functions\cite{gang}\cite{ccjpap} there is a trace with the
following Dirac matrices:
$$
G=  \frac {1}{2} (\gamma ^{+}+h\gamma ^{+}\gamma _{5}+
                \gamma ^{+}\gamma _{5}\mynot{s_{T}}).
\EQN{gamsum}
$$
The light-cone component, manifest in the overall factor of
$\gamma ^{+}$, provides the projection on the
leading twist part of the fragmentation.  \Eq{gamsum}
gives the Feynman rule for the cut eikonal vertex that
defines the fragmentation function.  The $\gamma ^{+}$ term gives the
unpolarized fragmentation.  The helicity dependent term
$h\gamma ^{+}\gamma _{5}$ will not contribute since we only measure two
particles in the final state; a helicity dependent asymmetry only
occurs with the three-particle correlation called
``handedness''.\cite{nacht}\cite{umconf}\cite{efremov}

The Feynman rules for fragmentation functions were written
down in \Ref{jcpdf}, with the polarization dependence given
in \Ref{gang}\Ref{ccjpap}.
The distribution of a single measured hadron $A$ of momentum
$p$ in the fragmentation of a quark is a function of the
longitudinal momentum fraction $z$ of the hadron and of the
transverse momentum $\kt$ of the quark relative to
the hadron.  It has the form\cite{ccjpap}
$$
\hat{D}_{A/\psi }(z,\kt)=
\int {dk^{-}\over (2\pi )^{4}} \Phi _{A/\psi }(k^{2},k\cdot p),
\EQN{phifun}
$$
where $k^{+}=p^{+}/z$, and $\Phi _{A/\psi }(k^{2},k\cdot p)$ is the inclusive
two--point function,
$$
\Phi _{A/\psi }(k^{2},p\cdot k)= \frac {1}{6}\int d^{4}x e^{ik\cdot x}
\Tr G \langle 0|\psi (x) a_{{\rm out}A}^{\dagger }(p)
                a_{{\rm out}A}(p) \bar\psi (0)|0\rangle ,
\EQN{twopt}
$$
with the Dirac matrix $G$ being defined in \Eq{gamsum}.  The
overall factor $1/6$ represents an average over quark color and spin.
This is represented diagrammatically in
\Fig{fthree}.
The factor $a_{{\rm out}A}^{\dagger }(p) a_{{\rm out}A}(p) $
is the number operator which counts how
many hadrons of type $A$ and momentum $p$ are in a state.
Note that the transverse momentum of the hadron relative to
the quark is $-z\kt$.

\figure{fthree}
\epsfxsize=3in
\centerline{\epsfbox{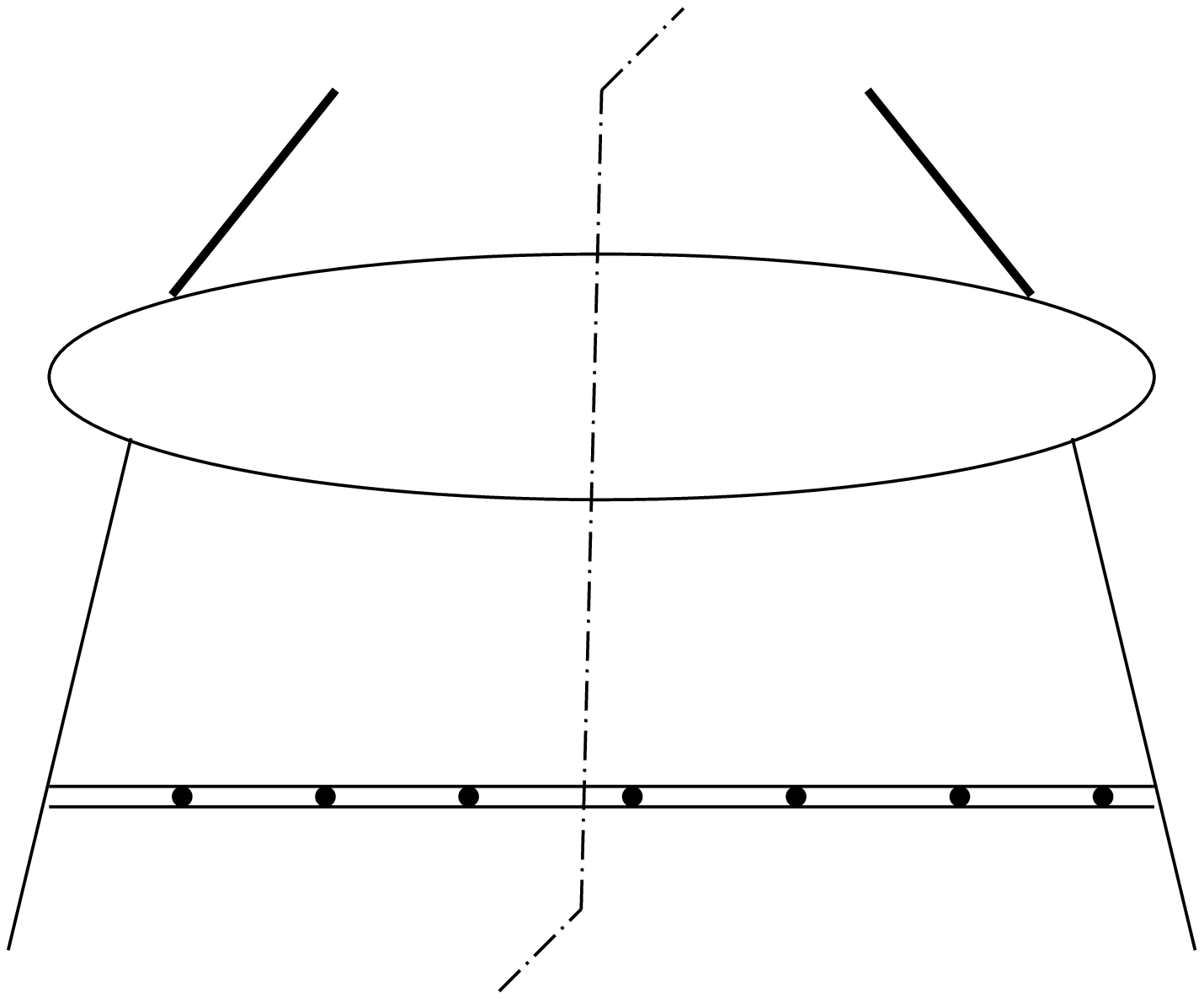}}
\caption{The fragmentation of a $\psi $ into hadron $A$.}
\endfigure

The usual decay function is obtained by integrating over
transverse momentum:
$$
D_{A/\psi }(z)=\int d^{2}\kt \hat{D}_{A/\psi }(z,\kt).
\EQN{daphi}
$$
It has no dependence on the quark spin $(h,s)$.
To interpret the fragmentation function
as a probability would require an extra factor, but
since we will be computing polarization asymmetries, this factor will
cancel between the numerator and the denominator, so we will ignore
it.  For further details the reader is referred to \Ref{jcpdf} and
\Ref{ccjpap}.

Exactly similar formulae apply to fragmentation to 2
measured particles.  We simply replace the number operator
$a_{{\rm out}A}^{\dagger }(p) a_{{\rm out}A}(p) $ by the corresponding
operator for two particles:
$$
a_{{\rm out}A_{1}}^{\dagger }(p_{1})
a_{{\rm out}A_{2}}^{\dagger }(p_{2}) a_{{\rm out}A_{2}}(p_{2})
a_{{\rm out}A_{1}}(p_{1}),
$$
so that the fragmentation function to two particles is
defined by
$$\eqalign{
\hat D_{H/a}(z,k_{\perp })
&\equiv \sum _{X}\int {dy^{-}d^{2}y_{\perp }\over 12(2\pi )^{3}}
e^{ik^{+}y^{-}-ik_{\perp }\cdot y_{\perp }}      \cr
&\times \Tr \gamma ^{+}
\langle 0|\psi _{a}(0,y^{-},y_{\perp })|A_{1}A_{2}X\rangle
\langle A_{1}A_{2}X|\bar{\psi }_{a}(0)|0\rangle  .
\cr}
\EQN{unpolpipi}
$$
The polarization-dependent part is\cite{gang}\cite{ccjpap}
$$\eqalign{
\Delta \hat{D}_{H/a}(z,k_{\perp },s_{\perp })
&\equiv \sum _{X}\int {dy^{-}d^{2}y_{\perp }\over 12(2\pi )^{3}}
e^{ik^{+}y^{-}-ik_{\perp }\cdot y_{\perp }}      \cr
&\times \Tr\gamma ^{+}\gamma _{5}\gamma _{\perp }\cdot s_{\perp }
\langle 0|\psi _{a}(0,y^{-},y_{\perp })|A_{1}A_{2}X\rangle
\langle A_{1}A_{2}X|\bar{\psi }_{a}(0)|0\rangle  .
\cr}
\EQN{jcdefa}
$$

\figure{ffour}
\epsfxsize=6in
\centerline{\epsfbox{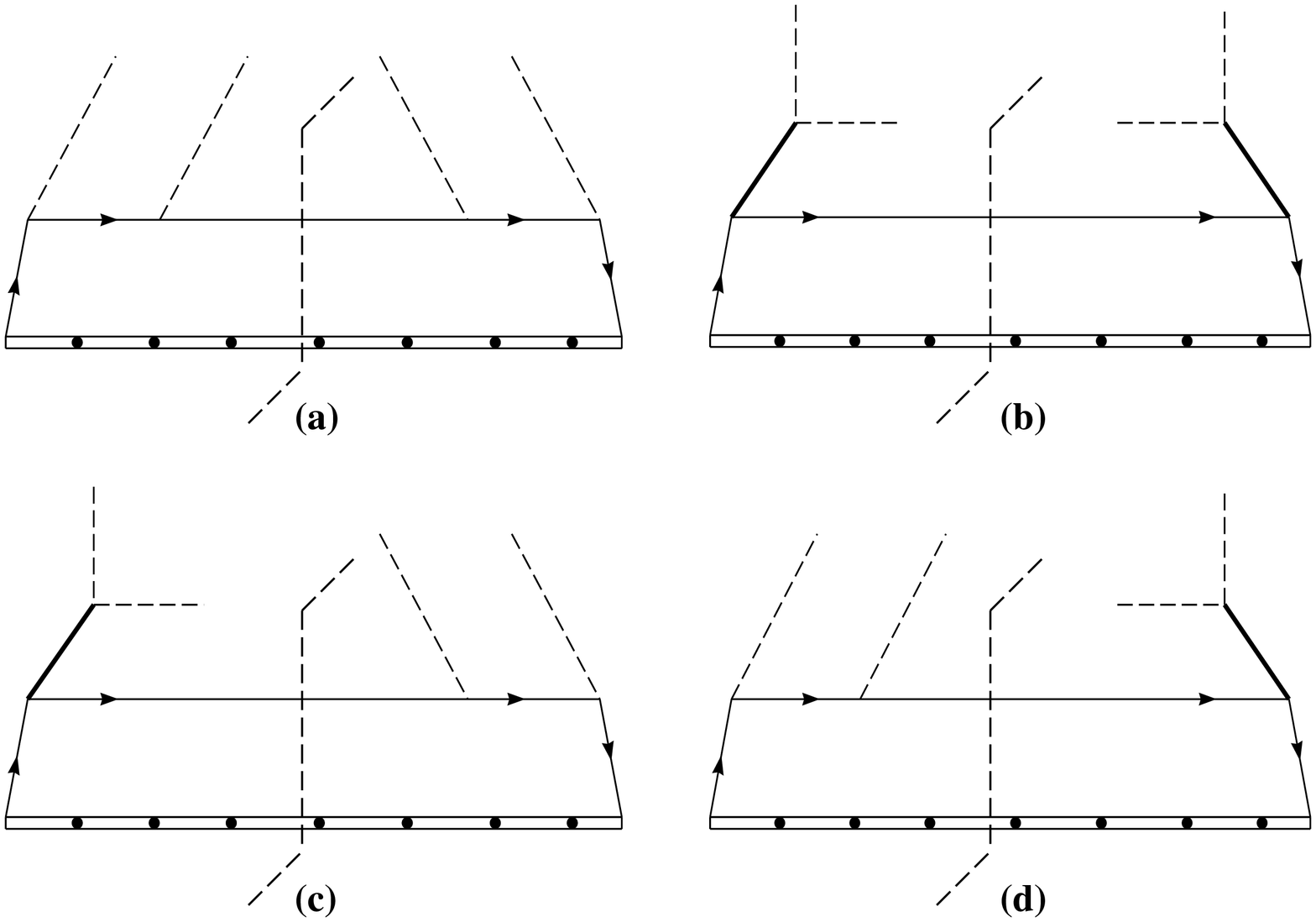}}
\caption{The four cut diagrams for the $u^{*}\to \pi ^{+}\pi ^{-} u$
fragmentation}
\endfigure

Ultimately we will integrate over $k_{\perp }$, the transverse momentum
of the quark relative to the measured hadrons, and in particular
this will imply an azimuthal average over $k_{\perp }$. In this
situation, to get transverse-spin dependence
will require that there be an
imaginary part in the amplitudes making up \Eq{jcdefa} (since the
factor of $\gamma _{5}$ will give a factor $i$).
In lowest order for the
fragmentation $q^{*}\to \pi \pi X$, the imaginary part will come from
the $\sigma $ propagator, as in \Fig{ffour}(c) and (d).

First, we will compute the
fragmentation for $q^{*}\to \sigma X$, followed by the decay
$\sigma \to \pi \pi $.
This will give us the total rate for $q^{*}\to \pi \pi X$ at the resonance.
Then we will compute the spin dependence from interference
graphs for $q^{*}\to \pi \pi q$, with a correct treatment of the imaginary part
of
the denominator of the $\sigma $ propagator

The momentum of the
fragmenting quark is denoted by $k$ and, depending upon the situation,
the momentum of either the $\sigma $ or $\pi \pi $ pair is represented by $p$.
The Lorentz frame chosen is that where the sigma particle
(or rather the combined momentum of the two pions of interest) has
zero transverse momentum.
The longitudinal momentum fraction of the two-pion system (or $\sigma $)
is given by $z\equiv p^{+}/k^{+}$.  Generally, these two-pion system variables
will be written in terms of individual pion variables,
$$
z=z_{1}+z_{2},    \qquad  p=p_{1}+p_{2},
\EQN{twopi}
$$
where the individual pion variables will be defined according to the
process considered.

\subsection{Decay Width of Sigma}

The imaginary part that we need for the spin-dependence
comes from the imaginary part of the denominator of
the $\sigma $ propagator.  At lowest order, this will be given by
the width of the $\sigma $.  Recent work has appeared where the random
phase approximation\cite{swidtha} and one--loop computations\cite{swidthb}
have been applied to the linear sigma model to
compute the width $\Gamma (\sigma \to \pi \pi )$.
However, since we are simply demonstrating the existence of a spin observable
in this paper, we shall not enter the quantitative detail of these references.

With the $\sigma {-}\pi {-}\pi $ coupling described by \Fig{ftwo}, the decay
width may be expressed as
$$
\Gamma (\sigma \to \pi \pi )={3\lambda ^{2}v^{2}\over 8\pi m_{\sigma }},
\EQN{sigwidth}
$$
where two thirds of the width is due to the decay to charged pions and
one third of the width is due to the decay to the neutral pion pair.
Using $m_{\sigma }=600\,$MeV, $m_{q}=300\,$MeV, and $v=92\,$MeV,
we find that $\Gamma (\sigma \to \pi \pi )=761\,$MeV.
The relations between the parameters which exist due to the
spontaneous symmetry breaking, allow us to write the width in
other forms:
$$
\Gamma (\sigma \rightarrow \pi \pi )=
{3\gee^{2}m_{\sigma }^{3}\over 32\pi m_{q}^{2}}
   = {3 m_{\sigma }^{3} \over 32 \pi  v^{2} }.
\EQN{symsigwid}
$$

\section{Fragmentation $q^{*}\rightarrow \sigma X$}

Our goal is to compute the $\pi \pi $ asymmetry
near the $\sigma $ pole, so we next find the total
production rate for fragmentation to $\sigma $.  This
rate will provide the normalization for the polarization asymmetries.
The distribution to be evaluated is represented by the
diagram in
\Fig{ffour}(b), but with the pion loop removed.
The momenta are defined such that $k$ represents the incoming
virtual quark momenta while $p_{3}$ and $p$ respectively represent the
outgoing quark and sigma momentum.  (The momenta $p_{1},p_{2}$ will
represent the pion momenta in $q^{*}\to q\pi \pi $.)
The $+$ momentum fraction carried away by the sigma is given by $z$.
Summing over final state quark spins and averaging over initial
state spins and colors yields the relation
$$
\eqalign{
\hat D_{\sigma /q}(z,k_{T})
=& 2\pi \gee^{2}\int {dk^{+}dk^{-}\over (2\pi )^{4}}
{\Tr[(\mynot{p}_{3}+m_{q})(\mynot{k}+m_{q})\gamma ^{+}(\mynot{k}+m_{q})]
\over 4(k^{2}-m_{q}^{2})^{2}}       \cr
&\qquad\qquad\qquad\qquad\qquad
\delta (k^{+}-{p^{+} /z})\delta ((k-p)^{2}-m_{q}^{2}),
\cr}
\EQN{sigqone}
$$
where the computation is being done in $d=4$ spacetime dimensions.
There will be no spin dependence for a single particle
fragmentation after we integrate over the transverse
momentum (or at least its azimuth).

The definition of the fragmentation function is that
we use a frame in which the $\sigma $ has zero transverse momentum,
${\bf p}_{T}=0$.  A simple calculation yields
$$
\hat{D}_{\sigma /q}(z,k_{T})= {\gee^{2} z^{2}\over 16\pi ^{3}}
   \ { m_{q}^{2}(2-z)^{2}+z^{2} k_{T}^{2}
     \over
       [m_{\sigma }^{2}(1-z)+m_{q}^{2}z^{2}+z^{2} k_{T}^{2}]^{2}
     }.
\EQN{sigqfiveb}
$$

Normally, fragmentation functions are defined with an integral
over $k_{T}$, the transverse momentum of the quark relative to the
measured hadrons.  These integrals have divergences at large $k_{T}$
which are removed by renormalization (or some other means, like a
cut off).  The dependence on the cut-off or on the
renormalization scale gives the usual Altarelli-Parisi evolution.
For our purposes, however, it will be convenient not to bother
with the integral.   In any event, our model is only appropriate
integrated over low $k_{T}$, where QCD is non-perturbative: The
fragmentation functions would then be suitable initial data for
normal perturbative QCD evolution.

Therefore we will not present the results integrated over $k_{T}$.

To obtain the resulting fragmentation function to $\pi ^{+}\pi ^{-}$,
we must insert the decay $\sigma \to \pi ^{+}\pi ^{-}$ into the quark
fragmentation derivation.  This is given by the sigma propagators
and the pion loop in \Fig{ffour}(b), which result in \Eq{sigqone}
being multiplied by a factor
$$
  \left|
     {i (-2i\lambda v) \over p^{2}-m_{\sigma }^{2}+im_{\sigma }\Gamma }
  \right|^{2}.
\EQN{extrafa}
$$
We are aware that the width of the sigma is in the neighborhood of its mass
(using $m_{\sigma }=600\,$MeV, $m_{q}=300\,$MeV, and $v=92\,$MeV),
but since our main
interest is to demonstrate a nonzero asymmetry, and its
order of magnitude, we just use the narrow width approximation,
$$
   \left|
      { i \over p^{2}-m_{\sigma }^{2}+im_{\sigma }\Gamma  }
   \right|^{2}
   \approx
   { \pi  \over m_{\sigma }\Gamma  }  \delta (p^{2}-m_{\sigma }^{2}).
\EQN{narwid}
$$

For the total rate, and in the narrow width approximation,
the graphs in
\Fig{ffour} other than (b) are unimportant, since they are less
singular at $p^{2} = m_{\sigma }^{2}$.
Hence the lowest order approximation for unpolarized
fragmentation to $\pi ^{+}\pi ^{-}$ pairs is
$$
\hat D_{\pi ^{+}\pi ^{-}/q} =
    { 32 \pi ^{2} \over 3}
    \hat D_{\sigma /q}(z_{1}+z_{2},\kt).
\EQN{calcpipi}
$$
The factor $32\pi ^{2}/3$ can be considered as $16\pi ^{2}$ times the
branching ratio, $2/3$, to $\pi ^{+}\pi ^{-}$.

\section{Polarized Partons and Two--Particle Fragmentation}

We now compute the dependence on transverse spin of the
fragmentation $q\to \pi ^{+}\pi ^{-}X$, using the definition \Eq{jcdefa}.
Since our asymmetry will be at the $\sigma $ pole, we define
the sigma momentum as the combined momentum of the two final
state pions.

We consider the decay of
an up quark which produces a charged pion pair.  At lowest
order, there are four
Feynman diagrams for the $u^{*}\to \pi ^{+}\pi ^{-} u$
fragmentation, and they are shown in \Fig{ffour}.  Topologically, this
decay is the same as the decay for the down quark,
$d^{*}\to \pi ^{+}\pi ^{-} d$.
Likewise, nonzero asymmetries are also expected from the
fragmentation to $\pi ^{\pm }{-}\pi ^{0}$ pairs.
The quark decays to $\pi ^{0}\pi ^{0}$
pairs will produce a smaller asymmetry, because the
azimuthal dependence is $\sin\phi $, and Bose symmetry of the
two pions therefore requires a zero in the spin asymmetry at
$z_{1}=z_{2}$.

\subsection{Getting the Polarization Dependence}

All the spin dependences for the lowest order
$u^{*}\to \pi ^{+}\pi ^{-} u$ are in the imaginary parts of the traces from
the cut diagrams.
Since the decay functions must be real, combining
the results from all of the relevant tree diagrams will result in the
cancellation of all of the spin contributions for the total decay
functions unless there is another factor present with an imaginary
part.  Such is the case when we are at a resonance.

Of the diagrams in \Fig{ffour} for $u^{*}\to \pi \pi u$,
the only ones which can contribute a transverse
spin dependence to the decay functions are (c) and (d).
They have interference between the continuum and a $\sigma $
propagator.
We next note that these diagrams each have one factor
of the $\sigma $-propagator, for which we can write
$$
\frac {i}{p^{2}-m_{\sigma }^{2}+im_{\sigma }\Gamma } =
PV \frac {i}{p^{2}-m_{\sigma }^{2}} + \pi  \delta (p^{2}-m_{\sigma }^{2}),
\EQN{prop}
$$
in the narrow width approximation.  The sum of the lowest order
decay function amplitudes represented by \Fig{ffour}(c) and~(d)
can then be expressed as
$$
  2 \Re\left[
        { a+ib \over p^{2}-m_{\sigma }^{2}- im_{\sigma }\Gamma  }
    \right].
\EQN{spindecomp}
$$
Here $a$ and $b$ are smooth functions of the kinematic
variables in the vicinity of the $\sigma $ pole.  Moreover, as we
will see, the spin dependence resides only in the $b$
coefficient.  The principal value in \Eq{prop} implies that
the spin-independent term gives a small contribution when
integrated over a neighborhood of the $\sigma $ pole, so that
\Eq{calcpipi} gives the unpolarized fragmentation.  Nevertheless, it
is \Eq{spindecomp} that gives the singular part of the spin
dependence:
$$
    2\pi  \delta ((p_{1}+p_{2})^{2}-m_{\sigma }^{2}) b.
\EQN{spindeppart}
$$

The decay function can then be expressed as the sum of an unpolarized
contribution and a spin asymmetric piece
$$
\hat {D}_{q^{*}\to \pi \pi q}(z_{1},z_{2}, s_{T}, \kt)=
\hat{D}_{q^{*}\to \pi \pi q}^{unpol}(z_{1},z_{2},\kt)
+\Delta \hat {D}_{q^{*}\to \pi \pi q}^{pol}(z_{1},z_{2},s_{T},\kt) .
\EQN{dparts}
$$
In the narrow width approximation, both the unpolarized and the
polarized term have a factor $\delta ((p_{1}+p_{2})^{2}-m_{\sigma }^{2})$, so
that the
ratio, which is the spin asymmetry, is well-defined:
$$
   \frac {\Delta \hat{D}_{q^{*}\to \pi \pi q}^{pol}(z_{1},z_{2},s_{T},\kt)}{
\frac {32\pi ^{2}}{3}\hat D_{\sigma /q}(z_{1}+z_{2},\kt)} .
\EQN{Spinasym}
$$
Here, we used \Eq{calcpipi} for the denominator.

\subsection{The Amplitude Computations}

We now compute the necessary terms from the diagrams in
\Fig{ffour} for the fragmentation of an up quark into
two charged pions.
We denote $\pi ^{+}$ momentum by $p_{2}$ and the $\pi ^{-}$
momentum by $p_{1}$,
and we use  a label ($a,b,c,d$) to indicate the
particular diagram in \Fig{ffour}.
$$
\eqalign{
  F_{c}= -2\pi {\gee}^{3}\lambda vC_{c}
  & {
      \Tr[(\mynot{p}_{3}+m_{q})
      (\mynot{k}+m_{q})G(\mynot{k}+m_{q})(\mynot{k}-\mynot{p}_{2}-m_{q})]
   \over
      (k^{2}-m_{q}^{2})^{2}[(k-p_{2})^{2}-m_{q}^{2}]
      [(p_{1}+p_{2})^{2}-m_{\sigma }^{2}+im_{\sigma }\Gamma ]
   }
\cr
   & \delta ((k-p_{1}-p_{2})^{2}-m_{q}^{2}) ,
}
\EQN{Fs}
$$
The $C_{c}$ factor carries the isospin factors and for
$u^{*}\to \pi ^{+}\pi ^{-} u$, $C_{c}=C_{d}=2$.
Graph (a) has no singularity at the $\sigma $ pole,
so that we do not need to calculate it for our approximation.
We have
already obtained the contribution of graph (b), in \Eq{calcpipi}.
Notice that $F_{d}$ is the complex conjugate of
$F_{c}$, so that we do not need to calculate it explicitly.

Therefore, we obtain the spin dependent portion of the decay
function from
$$
   \Delta \hat{D}(z_{1},z_{2},{\bf s}_{T },\kt) =
   [\int {dk^{-}\over (2\pi )^{4}} (F_{c}+F_{d})]_{(p_{1}+p_{2})^{2}=m_{\sigma
}^{2}} =
   \Re [2 \int {dk^{-}\over (2\pi )^{4}} F_{c}]_{(p_{1}+p_{2})^{2}=m_{\sigma
}^{2}} ,
\EQN{fulldb}
$$
If we equate $F_{c}+F_{d}$ from \Eq{Fs} with \Eq{spindecomp},
we find the coefficient $b$ for $u^{*}\rightarrow \pi ^{+}\pi ^{-}u$
takes the value
$$\eqalign{
   b=&
   {
      -2\pi \lambda v\gee^{3} C_{c} \delta (p_{3}^{2}-m_{q}^{2})
   \over
       2(p_{1}\cdot p_{3})[k^{2}-m_{q}^{2}]^{2}
   }
\cr
   & \left\lbrace
   4[({p^{+}\over z})\veps (p_{1},p_{3},s,p)-hm_{q}\veps (p_{1},p_{3},p,+)
   +(s\cdot k)\veps (p_{1},p_{3},k,+)]     \right.
\cr
   &\left. \qquad\qquad\qquad\qquad\qquad\qquad\qquad\qquad\qquad\qquad
   -\half (p^{2}+2p\cdot p_{3})\veps (p_{1},p_{3},s,+)\right\rbrace ,
}
\EQN{bcd}
$$
where $p_{3}=k-p_{1}-p_{2}$ is the momentum of the on--shell final state
quark, and $p=p_{1}+p_{2}$. We define $\epsilon (p,q,r,s)\equiv \epsilon
_{\kappa \lambda \mu \nu } p^{\kappa }q^{\lambda }r^{\mu }s^{\nu }$,
with the
Levi-Civita symbol $\epsilon _{\kappa \lambda \mu \nu }$ obeying the following
relations in light
cone coordinates (LCC),
$$\eqalign{
\veps ^{\mu \nu \pm \pm }=& 0 \cr
\veps ^{\mu \nu \pm \mp }=& \mp \veps ^{\mu \nu 03}
\qquad\hbox{(using $1\over\sqrt {2}$ normalization per LCC index)}\cr
\veps (p,k,s,0)=& -(\vec{p}\times \vec{k})\cdot \vec{s} ,
\cr}
\EQN{levic}
$$
where the convention is set by $\veps _{0123}=+1$.

Since we will use the fragmentation function after integrating
over $\kt$, we will average over the azimuth of $\kt$. After this
average, the helicity term vanishes because the formulas no
longer maintain a sufficient number of Lorentz vectors to keep
the contraction with the Levi--Civita tensor asymmetric in its
indices. We also apply the narrow width approximation provided by \Eq{prop}.
We obtain
\def\ksqpmsq{{k_{T }^{2}+m_{q}^{2}}}
$$\eqalign{
   \Delta \hat{D}
   =& \delta ((p_{1}+p_{2})^{2}-m_{\sigma }^{2})
   {\gee^{3} \lambda v \over 2\pi ^{2}}
   {\hat{\bf z}\cdot (\vec{\bf s}_{T }\times \vec{\bf p}_{1T})
   \over
     2p_{1T}^{2}
   }
   {z_{\sigma }^{2} (1-z_{\sigma })
   \over
   [(k_{T }^{2}+m_{q}^{2})z_{\sigma }^{2}+m_{\sigma }^{2}(1-z_{\sigma })]^{2}
   }
\cr
   \Bigg\lbrack    m_{\sigma }^{2} z_{2}-
   &
   {
      (\ksqpmsq )(z_{1}^{2}z_{2}m_{\sigma }^{2}-2z_{1}z_{\sigma
}^{2}p_{1T}^{2})+
      (1-z_{\sigma })[z_{2}(1-z_{\sigma })m_{\sigma }^{2}-
              4z_{1}z_{\sigma } m_{q}^{2}]p_{1T}^{2}
   \over
      \sqrt {[(1-z_{\sigma })^{2}p_{1T}^{2}+z_{1}^{2}(\ksqpmsq )]^{2}-
               4z_{1}^{2}(1-z_{\sigma })^{2}p_{1T}^{2}k_{T}^{2}}
    }
\Bigg\rbrack    ,
}
\EQN{deldhat}
$$
where $z_{\sigma }=z_{1}+z_{2}$ and $\hat{\bf z}=(0,0,1)$ in rectangular
coordinates.

The main demonstration of this paper is complete.  Using the linear sigma
model, we have achieved a nonvanishing asymmetry in the transverse polarization
of a quark at leading twist.  The nonvanishing component in \Eq{deldhat}
was maintained through the interference of the continuum production of $\pi \pi
$
pairs with the sigma resonance.  Furthermore, it is apparent that the
existence of \Eq{deldhat} is dependent upon a nonzero value for the expectation
value of the sigma field, i.e., it is the broken chiral symmetry which
permits this asymmetry to exist.   Consequently, this asymmetry can
(theoretically) probe the chiral nature of QCD.
In the limit that the bare quark mass is
zero, nonvanishing polarization effects remain and without
singularity as long as $z$ is not at its endpoints.

The asymmetry at the sigma resonance may be estimated by comparing the
spin dependence of $q^{*}\to q\pi \pi $ given by \Eq{deldhat} with
the unpolarized fragmentation given by
\Eq{calcpipi}.  One immediate consequence is that the
the correlation of the pions' direction with the
transverse quark spin is given by the cross product
$(\vec{\bf s}_{T}\times \vec{\bf p}_{1T})$.  This is in accordance with the
general theory.\cite{gang}
So, as the pion momentum
vector rotates about
the quark (jet) axis, the asymmetry exhibits a sinusoidal rise and fall
as it moves with respect to the transverse spin direction of the quark.

We have used the perturbative approximations, low order graphs and the
narrow width approximation ($\Gamma \ll m_{\sigma }$), but these are
not good approximations.  Moreover, the use of both quark and hadron
degrees of freedom as we have done is a crude model.  So the following
quantitative calculations must be only considered very rough
estimates.  Nevertheless, they do indicate that the spin-dependence is
as large as it can be.

In \Tbl{ktasym}, we list some values of the asymmetry: the ratio
of \Eq{deldhat} to \Eq{calcpipi}.  We have replaced the triple
product $(\vec{\bf s}_{T}\times \vec{\bf p}_{1T})$ by $|p_{1T}|$, so that the
numbers represent the amplitude of the sinusoidal dependence of
the pion production on the azimuthal angle about the jet axis.

\midtable{ktasym}
\caption{The asymmetry of \Eq{Spinasym} is tabulated for
  various values of $k_{T}$, $z_{1}$ and $z_{2}$ using
  $s_{T}\times p_{1T} \to  |s_{T}||p_{1T}|$.
  We use the values
  $m_{\sigma }=600\,$MeV, $m_{q}=300\,$MeV, $v=92\,$MeV.
}
\singlespaced
\ruledtable
\multispan2
\hfil THE TRANSVERSE SPIN ASYMMETRY \hfil\CR
$z_{1}=0.4$, $z_{2}=0.3$ \CR
$z_{\sigma } k_{T}$ (MeV)      |  Asymmetry \CR
0                |     1.31   \cr
200              |     1.21   \cr
400              |     0.89   \cr
600              |     0.58   \cr
1200             |     0.20   \CR
$z_{1}=0.2$, $z_{2}=0.15$            \CR
$z_{\sigma } k_{T}$ (MeV)      |  Asymmetry \CR
0                |     0.31   \cr
200              |     0.31   \cr
400              |     0.34   \cr
600              |     0.39   \cr
1200             |     0.54
\endruledtable
\endtable

Notice that the numbers are large.  One might worry that the
asymmetries go above $100\%$, and that within a lowest order
perturbative calculation (from the graphs of \Fig{ffour}) this
would correspond to a negative cross section.  However, we are
working in the neighborhood of the pole of the $\sigma $.  Therefore,
it was essential to use a dressed propagator for the $\sigma $ lines,
to get a suitable imaginary part and resonance width.  This means
that our model is not totally self-consistent: We have performed
a selective resummation of graphs.  We have further made a narrow
width approximation on the $\sigma $ pole: This is obviously far from
perfect for physical values of the couplings of our model.

Our approximations are valid in the weak coupling limit,
$\gee, \lambda  \ll 1$, whence the asymmetry is of order
$\gee \sqrt \lambda $,
which is then much less than one.  The true strong interactions
are, of course, strong, and so obtaining an asymmetry above unity
from our calculation is not impossible.  All it indicates is that
the effect we are calculating suffers from no particular
suppression.  Hence we can expect substantial analyzing power for
quark transverse spin from measurements of the azimuthal
dependence of pion pairs.

Note that we have used $z_{\sigma }k_{T}$, rather than $k_{T}$, as the
transverse momentum variable, since this represents the
transverse momentum of the pion pair relative to the jet.

When $z_{\sigma }k_{T}$ gets large, the asymmetry decreases.  This is
expected, since the quark is then far off-shell, and the graphs
approach their values with zero quark mass.  When the quark mass
is zero, there is exact helicity conservation along the quark
lines, and hence there is no transverse-spin asymmetry.  It can
be easily checked from \Eq{deldhat} and \Eq{calcpipi}
that our calculated asymmetry is proportional to $1/k_{T}^{2}$ at large
$k_{T}$.

At small $z_{\sigma }$, the calculated asymmetry appears to get smaller,
with a broader distribution in $z_{\sigma } k_{T}$.  This may be a good
prediction.  But our model should not be reliable at small
$z$.  A reasonable prejudice is only that the spin dependence
should decrease at small $z$, since there one expects hadron
production to be independent of the flavor and spin state of the
initiating quark: The associated pomeron and gluon physics are
not in our model.

\section{Conclusion}

Using the Linear Sigma Model to describe the fragmentation of a polarized
quark has demonstrated the existence of a nonvanishing asymmetry from which
polarization information can be obtained.  Although the model
should only be considered to give crude qualitative information,
the large asymmetries we calculate do show that the spin
correlation can be  completely unsuppressed.

This supports the value of doing experiments to measure the
asymmetry experimentally.  It can be used, for example, as a
method of obtaining the transverse spin dependence of quarks in
a transversely polarized proton.\cite{gang}

It is clearly important to find better models for the
fragmentation that include spin effects.  Obviously, real
fragmentation of a quark results in many more than two pions.
This can easily dilute the effect of the physics we have
modeled.  Nevertheless, the number of pion pairs is fairly low if we
restrict our attention to large $z$ and fairly low invariant
mass.

\nosechead{Acknowledgements}

The authors wish to thank J.~Botts and M.~Strikman, in
particular, for useful conversations.
This work was funded in part by
DOE grant DE-FG02-90ER-40577 and TNRLC grant RGFY9240.

\vfill\eject

\nosechead{References}
\ListReferences
\vfill\eject
\vfill\eject
\bye